# Self-consistent modelling of nonlinear dynamic ESM microscopy in mixed ionic-electronic conductors

O.V. Varenyk[1], M.V Silibin[2], D.A Kiselev[2,3], E.A. Eliseev[4], S.V. Kalinin[5*], A.N. Morozovska[1†]

[1] Institute of Physics, National Academy of Sciences of Ukraine, 46, pr. Nauki, 03028 Kyiv, Ukraine

[2] National Research University of Electronic Technology "MIET", 124498 Moscow, Russia

[3] National University of Science and Technology "MISiS", 119049 Moscow, Leninskiy pr. 4, Russia

[4] Institute for Problems of Materials Science, NAS of Ukraine, Krjijanovskogo 3, 03142 Kyiv, Ukraine

[5] The Center for Nanophase Materials Sciences, Oak Ridge National Laboratory, Oak Ridge, TN 37831

## Abstract

Dynamic Electrochemical Strain Microscopy (ESM) response of mixed ionic-electronic conductors is analysed in the framework of the Thomas–Fermi screening theory and Vegard law with accounting of the steric effects. The emergence of dynamic charge waves and nonlinear deformation of the surface as result of applying probing voltage is numerically explored. 2D maps of the strain and concentration distribution across the mixed ionic-electronic conductor and bias-induced surface displacements for ESM microscopy were calculated. Obtained numerical results can be of applied to quantify ESM response of Li-based solid electroytes, materials with resistive switching and electroactive ferroelectric polymers, which are of potential interest for flexible and high-density non-volatile memory devices

[*] Corresponding author: e-mail sergei2@ornl.gov
[†] Corresponding author: e-mail anna.n.morozovska@gmail.com

# 1. Introduction

Nanoscale properties of mixed conductors such as Li-based solid electroytes with mobile ions and electrons as free carriers, materials with memristive resistive switching such as manganites with mobile oxygen vacancies [1], electroactive ferroelectric polymers are intriguing and important [2, 3, 4, 5]. In particular synthetic electroactive polymers are of great importance for a number of research fields including biocompatible tissue engineering and organic electronics [6, 7, 8]. With increasing necessities for improving batteries performances, flexible and high-density non-volatile memory devices, the abovementioned mixed conductors being suggested as prospective candidates of technological interest.

Electromechanical properties of mixed conductors are of particular interest. Coupling between electrical and mechanical phenomena is one of the fundamental processes in nature manifested in physical objects ranging from ferroelectrics to biometric and biological systems [9]. Electromechanics refers to a broad class of phenomena in which mechanical deformation is induced by an external electric field, or, conversely, electric charge separation is generated by the application of an external force. In most materials, electromechanical activity is directly related to structure and functionality and thus is important not only for applications, but also for materials characterization. In polar compounds, local piezoelectric properties are strongly affected by polarizability, structural defects and mechanical properties. Obviously, progress in fundamental studies and technological applications of these materials depend on the ability of testing their structural and functional properties at nanoscale.

The progress in understanding nanoscale electromehcnical phenomena had been achieved with the emergence of voltage modulated scanning probe microscopy (SPM) techniques such as Piezoresponse Force Microscopy [10,11, 12] (PFM) and Electrochemical Strain Microscopy (ESM) [13, 14, 15, 16, 17, 18, 19]. In PFM the tip creates electric field in the material and further detects piezoelectric surface deformation. In ESM, the biased tip acts as a moving, electrocatalytically active probe exploring local electrochemical activity. The probe concentrates an electric field in a nanometer-scale volume of material. The electric field altered the local electrochemical potential of the lithium ions on the surface, causing them to intercalate or de-intercalate. This changes the local concentration of mobile ions by migration (field-driven) and diffusion (concentration gradient-driven) mechanisms, and also changed the lattice volume under the tip.

The associated changes in molar volume [20, 21] result in local electrochemical strains, resulting from the changes in ion concentration and associated dynamic surface deformation is detected by SPM at the 2–5 pm level. The tip–electrolyte contact can be described as a harmonic oscillator, the resonant frequency of which is determined primarily by the Young's modulus of the contact area between the tip and electrolyte. Using a lock-in technique, the resonant amplitude of the surface displacement, measured in nano- or picometers, can be detected by the SPM tip, yielding information about the local bias-induced lithium concentration changes and thus lithium transport. ESM can detect volume changes corresponding to complete lithiation and delithiation on a single atomic layer, or ~5-10% changes in lithium concentration within the ~20-nm region [13-19]. Electrochemical reactions detected by ESM play an important role on operational mechanisms in memristive behavior such as electroforming and subsequent resistive switching [22, 23]. The localization of ESM signal at interfaces indicates that the latter are responsible for mechanical stability and irreversible capacity loss, suggesting possible strategies for optimization of various electrochemically active materials [24].

Available theoretical models of mixed conductors ESM response are mostly linear [25, 26, 27, 28]. The linear models proved that the coupling between ionic redistribution and Vegard strains give rise to the ESM response in these materials and describe adequately it frequency spectrum, but the observed ferroelectric-like shape of hysteresis loops remained mainly unexplained. Actually, thermodynamics of electromechanically coupled mixed ionic-electronic conductors is defined by Vegard strains and flexoelectric effect [25]. One-dimensional [26] and two-dimensional analytical models [27] of linearized diffusion kinetics in ESM, and 2D analytical model that considers linearized drift-diffusion kinetics [28] have been evolved and result to the elliptic loop shape with "coercive" voltage (in fact defined from the condition of zero response) determined by applied bias.

In contrast this study is the first self-consistent 2D-modeling of the local mechano-electro-chemical response of solid electrolytes utilizing kinetic theory with accounting of the steric effects for ions (or vacancies) [29, 30] and, thus, including the most common form of the nonlinearity inherent to the system. To get further insight into the mechanisms of ESM image formation we evolved a 2D analytical model of nonlinear drift-diffusion kinetics in ESM. Corresponding numerical finite element (FE) modeling for different frequencies and bias voltage amplitudes was performed for 2D axially symmetric geometry. The novelty of the obtain results is that they are focused on the nonlinear mechanisms of ESM. Obtained 2D maps provide concentration and strain distribution with accounting of the steric effects for donors, Vegard mechanism and electrostriction. In numerical modelling we assume the decoupling approximation and ignore contributions of strain in the ionic transport. The surface of the sample is assumed to be perfectly smooth without any irregularities or roughness.

## 2. The problem statement and basic equations

In the case of tip axial symmetry and homogeneous mixed ionic-electronic conductor all physical quantities depend only on the distances z from the tip-surface interface and polar radius *r* (**Fig.1**). Mobile positively charged point defects, oxygen vacancies or cations, further considered as *donors* for the free electrons are inherent to the film.

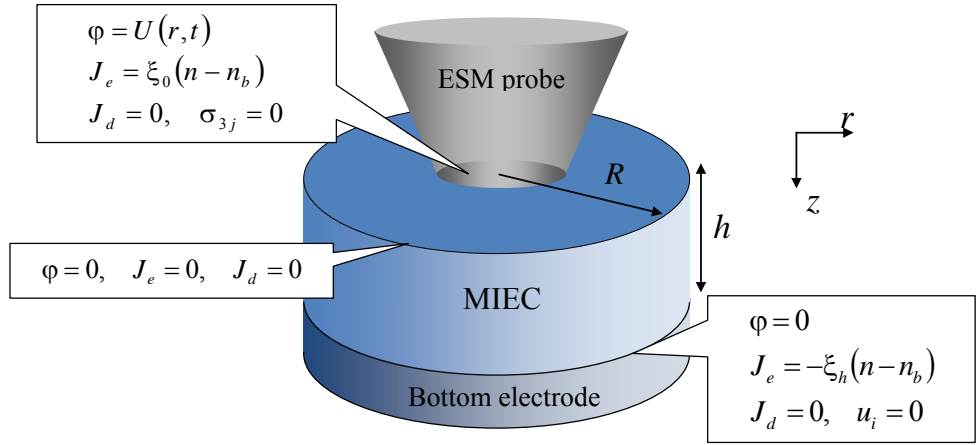

**Figure 1.** Typical geometry for ESM study of mixed ionic-electronic conductor (MIEC). Electrical and mechanical boundary conditions are labeled.

Mobile charge carriers redistribution creates the internal electric field that components $E_z = -\partial\varphi/\partial z$ and $E_r = -\partial\varphi/\partial r$ are defined by electric potential $\varphi$. The potential can be determined self-consistently from the Poisson equation that form in cylindrical coordinates is:

$$\varepsilon_0\varepsilon\left(\frac{\partial^2\varphi}{\partial r^2} + \frac{1}{r}\frac{\partial\varphi}{\partial r} + \frac{\partial^2\varphi}{\partial z^2}\right) = -e\left(Z_d N_d^+(\varphi) - n(\varphi)\right) \tag{1}$$

Here $\varepsilon_0 = 8.85 \times 10^{-12}$ F/m the dielectric permittivity of vacuum; $\varepsilon$ is a dielectric permittivity of MIEC, that is regarded isotropic, electron density is $n$, donor concentration is $N_d^+$, $e = 1.6 \times 10^{-19}$ C the electron charge, $Z_d$ is the donor charge that is equal to zero for uncharged vacancies or isovalent impurities. Electric potential satisfy fixed boundary conditions at the electrodes, $\varphi|_{z=0} = 0$, $\varphi|_{z=h} = U(r,t)$, $\varphi|_{r=R} = 0$, which correspond to the electroded MIEC film of thickness $h$. A periodic voltage $U$ is applied to the top electrode. Further we use its Gaussian form, $U(r,t) = U_0 \exp(-r^2/r_0^2) \sin(\omega t)$ and regard that the tip lateral size $r_0$ is much smaller that the size of the computation cell $R$, i.e. $r_0 \ll R$.

Continuity equation for donor concentration $N_d^+$ is:

$$\frac{\partial N_d^+}{\partial t} + \frac{1}{eZ_d}\left(\frac{1}{r}\frac{\partial (rJ_r^d)}{\partial r} + \frac{\partial J_z^d}{\partial z}\right) = 0, \qquad (2)$$

The donor current $J_d$ radial and normal components are proportional to the gradients of the carrier electrochemical potentials levels $\zeta_d$ as $J_r^d = -eZ_d \eta_d N_d^+ (\partial \zeta_d / \partial r)$ and $J_z^d = -eZ_d \eta_d N_d^+ (\partial \zeta_d / \partial z)$ respectively, where $\eta_d$ is the ions/vacancies mobility coefficient that is regarded constant. The boundary conditions for the donors are ion-blocking $J_d|_{z=0} = 0$; $J_d|_{z=h} = 0$; $J_d|_{r=R} = 0$

The electrochemical potential level $\zeta_d$ is defined as [31]:

$$\zeta_d = -E_d - W_{ij}^d \sigma_{ij} + eZ_d \varphi + k_B T \ln\left(\frac{N_d^+}{N_d^0 - N_d^+}\right). \qquad (3)$$

Here $E_d$ is the donor level, elastic stress tensor is $\sigma_{ij}$, $T$ is the absolute temperature, $k_B$ is a Boltzmann constant, $W_{ij}^d$ is the Vegard strain tensor (other name for it is elastic dipole). Hereinafter the Vegard tensor is regarded diagonal, i.e. $W_{ij}^d = W\delta_{ij}$ ($\delta_{ij}$ is delta Kroneker symbol). For the case of its diagonality $W_{ij}^d \sigma_{ij} \equiv W\sigma$, and the first stress invariant is introduced as $\sigma = \sigma_{zz} + \sigma_{rr} + \sigma_{\varphi\varphi}$. The absolute values of $W$ for ABO$_3$ compounds can be estimated as $|W| \propto (1 - 50)$ Å$^3$ from the refs [32, 33]. The maximal possible concentration of donors is $N_d^0$; it takes into account steric effects [29, 30]. For numerical estimates $N_k^0 \equiv a^{-3}$, where $a^3$ is the maximal volume allowed per donor centre. The steric effect limits the donor accumulation in the vicinity of film surfaces.

Continuity equation for electrons is:

$$\frac{\partial n}{\partial t} - \frac{1}{e}\left(\frac{1}{r}\frac{\partial (rJ_r^e)}{\partial r} + \frac{\partial J_z^e}{\partial z}\right) = 0 \qquad (4)$$

The electron current radial and normal components are $J_r^e = e\eta_e n (\partial \zeta_e/\partial r)$ and $J_z^e = e\eta_e n (\partial \zeta_e/\partial z)$ respectively, $\eta_e$ is the electron mobility coefficient, $\zeta_e$ is the electro-chemical potential. The boundary conditions for the electrons are taken in the linearized Chang-Jaffe (CJ) [34] form, $\left(J_z^e - \xi_0(n-n_b)\right)\big|_{z=0} = 0$, $\left(J_z^e + \xi_h(n-n_b)\right)\big|_{z=h} = 0$, $J_e\big|_{r=R} = 0$, where $\xi_{0,h}$ are positive rate constant related with the surface recombination velocity. CJ condition contains the continuous transition from the "open" electrode ($\xi_{0,h} \to \infty \Rightarrow n = n_0$) to the interface limited kinetics ($0 < \xi_{0,h} < \infty$) and "completely blocking" electrode ($\xi_{0,h} = 0$).

Continuous approximation for the concentration of the electrons in the conduction band is consistent with the following expression for electro-chemical potential [31]:

$$\zeta_e \approx E_C + k_B T\, F_{1/2}^{-1}\left(\frac{n(\varphi)}{N_C}\right) - e\varphi. \qquad (5)$$

The electro-chemical potential $\zeta_e$ tends to the Fermi energy level $E_F$ in equilibrium, $E_C$ is the bottom of conductive band, $F_{1/2}^{-1}$ is the function inverse to the Fermi integral $F_{1/2}(\xi) = \frac{2}{\sqrt{\pi}} \int_0^\infty \frac{\sqrt{\zeta}\, d\zeta}{1+\exp(\zeta-\xi)}$; effective density of states in the conductive band $N_C = \left(\frac{m_n k_B T}{2\pi\hbar^2}\right)^{3/2}$, electron effective mass is $m_n$. Electron density can be calculated from Eq.(5) as $n = N_C F_{1/2}\left((e\varphi + \zeta_e - E_C)/k_B T\right)$. Note, that the deformation potential effect, being the analog of the Vegard effect for electrons, is neglected in Eq.(5) for the sake of simplicity.

Using the dependencies for concentration of donors, $N_d^+ = N_d^0 f\left(-E_d - W\sigma + eZ_d\varphi - \zeta_d\right)$, ($f(x) = (1+\exp(x/k_B T))^{-1}$ is the Fermi-Dirac distribution function) and electrons, $n = N_C F_{1/2}\left((e\varphi + \zeta_e - E_C)/k_B T\right)$, on the electrochemical potentials $\zeta_{d,e}$, one can express the potentials as the functions of donor and electron chemical potentials $\mu_d = eZ_d\varphi - \zeta_d - W\sigma$ and $\mu_e = e\varphi + \zeta_e$ in the following way $N_d^+ = N_d^0 f(\mu_d - E_d)$ and $n = N_C F_{1/2}((\mu_e - E_C)/k_B T)$. Then coupled Eqs. (2) and (4) have the following form:

$$\frac{\partial f(\mu_d - E_d)}{\partial t} - \frac{1}{r}\frac{\partial}{\partial r}\left(r\eta_d f(\mu_d - E_d)\frac{\partial(eZ_d\varphi - \mu_d - W\sigma)}{\partial r}\right) - \frac{\partial}{\partial z}\left(\eta_d f(\mu_d - E_d)\frac{\partial(eZ_d\varphi - \mu_d - W\sigma)}{\partial z}\right) = 0 \quad (6)$$

$$\frac{\partial}{\partial t}F_{1/2}\left(\frac{\mu_e - E_C}{k_B T}\right) - \frac{1}{r}\frac{\partial}{\partial r}\left(r\eta_e F_{1/2}\left(\frac{\mu_e - E_C}{k_B T}\right)\frac{\partial(\mu_e - e\varphi)}{\partial r}\right) - \frac{\partial}{\partial z}\left(\eta_e F_{1/2}\left(\frac{\mu_e - E_C}{k_B T}\right)\frac{\partial(\mu_e - e\varphi)}{\partial z}\right) = 0 \quad (7)$$

The electrochemical strain can be further introduced as follows. Generalized Hooke's law for a chemically active elastic solid media relates the concentration deviation from the average $\delta N_d^+(\mathbf{r},t) = (N_d^+(\mathbf{r},t) - \overline{N}_d^+)$, mechanical stress tensor $\sigma_{ij}$, and elastic strain $u_{ij}$ through the equation

$$u_{ij} = W_{ij}\,\delta N_d^+ + s_{ijkl}\sigma_{kl} + \varepsilon_0^2 Q_{ijkl}\chi_{km}\chi_{lp}E_m E_p. \quad (8)$$

Here $s_{ijkl}$ is the tensor of elastic compliances, $Q_{ijkl}$ is electrostriction tensor, electric polarization $P_k = \varepsilon_0 \chi_{km} E_m$ for the considered linear dielectrics, and $W_{ij}$ is the Vegard expansion tensor.

The typical intrinsic resonance frequencies of material are in the GHz range, which is well above the practically important limits of ion dynamics and AFM-based detection of localized mechanical vibrations. This allows using quasi-static approximation for modeling of mechanical phenomena, namely we solve the general equation of mechanical equilibrium $\partial \sigma_{ij}/\partial x_j = 0$ in the quasi-static case. This leads to the equation for mechanical displacement vector $u_i$ inside of the film:

$$c_{ijkl}\left(\frac{\partial^2 u_k}{\partial x_j \partial x_l} - W_{kl}\frac{\partial \delta N_d^+}{\partial x_j} - \varepsilon_0^2 Q_{klst}\chi_{sm}\chi_{tp}\frac{\partial(E_m E_p)}{\partial x_j}\right) = 0. \quad (9)$$

Here $c_{ijkl}$ is the tensor of elastic stiffness. The boundary condition on the free surface of the film ($z = 0$) is $\sigma_{3j}(z = 0, t) = 0$. The surface $z = h$ is clamped to a rigid substrate and thus here $u_k\big|_{z=h} = 0$.

## 3. Electrostriction contribution to the ESM response in 1D case

Below we estimate the electrostriction contribution to the ESM response in one-dimensional (1D) approximation, and, for bulk polarization only. 2D case will be explored separately. To estimate, we consider several important components of MIEC film ESM response in 1D case: the Vegard one, electrostriction and their superposition. For these contributions consideration 1D approximation can be justified for the case when Debye length $h_d$ of MIEC is much smaller then the tip-surface contact radius $r_0$. For the case of the mixed ionic-semiconductor film of thickness $h$ placed in a planar capacitor under the application of *ac* and *dc* voltages superposition, $V_{dc} + V_{ac}\sin(\omega t)$, we calculated all abovementioned contributions to the displacement of the film surface as explained in **Appendix A** of Supplementary Materials.[35] Vegard contribution is proportional to the integral, $W\int_0^h \delta N_d^+ dz$, that is identically zero for the case of ion-blocking electrodes. Electrostriction contribution is proportional to the integral of polarization squire, $u_3(t) = Q_{33}\int_0^h P_3^2(z,t)dz$, and the polarization is approximated as $P(z,t) \approx P_W(z,t) + \varepsilon_0\chi(E_3^{dc}(z) + E_3^{ac}\sin(\omega t))$. Here $P_W(z,t) = eh_d\delta N_d(z,t)$ is the electric analog of the Vegard elastic dipole, $\chi$ is a static relative susceptibility. $E_{dc}$ is the slowly changing component of the electric field induced by tip bias, whereas $E_{ac}$ is the fast component. We note that for typical ionic systems the mobilities are sufficiently slow so that at excitation frequencies $\omega$ of about 100 kHz the $E_{ac}$ component can be evaluated as that for dielectric, neglecting ionic motion. For the cases of linear Debye screening (or no screening) of *ac* component and abrupt junction approximation for the space-charge density induced by *dc* component, the total mechanical displacement becomes:

$$u_3^Q(t) = u_3^0 + u_3^\omega \sin(\omega t) + u_3^{2\omega}\cos(2\omega t) \tag{10}$$

The "linear" coefficient $u_3^\omega = 2Q_{33}\varepsilon_0\chi P_W V_{ac} + Q_{33}(\varepsilon_0\chi)^2(V_{dc}V_{ac})/h$. Now we can estimate the relative contributions of electrostriction and Vegard effect to the linear dynamic piezoelectric coefficient, $d_{33}^\omega = du_3^\omega/dV_{ac} = 2Q_{33}\varepsilon_0\chi P_W + Q_{33}(\varepsilon_0\chi)^2(V_{dc}/h)$, using typical values of parameters $Q_{33}$=0.05 m$^4$/C$^2$, $P_W$=8×10$^{-3}$ C/m$^2$ ($e$=1.6×10$^{-19}$ C, $h_d$=5×10$^{-9}$m, $\delta N_d$=10$^{25}$m$^{-3}$), $\chi$~4, $\varepsilon_0 = 8.85 \times 10^{-12}$ F/m, $V_{dc}$=5 V, $h$~10 nm. The estimation gives $2Q_{33}\varepsilon_0\chi P_W$ ~ 0.03 pm/V and $Q_{33}(\varepsilon_0\chi)^2(V_{dc}/h)$~0.03 pm/V.

So in 1D-approximation both contributions proportional to the electrostriction coefficient appears much smaller than typical experimental detection limits ~1 pm/V. This allows us to concentrate all further attention on the Vegard strain contribution into ESM response at low voltages <1 V.

## 4. Results and discussion

All calculations were performed in the COMSOL multiphysics package using the "PDE" and "Solid Mechanics" modules with parameters listed in the **Appendix B** of Supplementary Materials, Table B1.[35] Numerical solution of Eqs.(6)-(7) was performed in the dimensionless variables listed in the appendix. For brevity we introduce only the main dimensionless variables and parameters in the main text, namely dimensionless cylindrical coordinates and thickness $\tilde{z} = z/L_D$, $\tilde{r} = r/L_D$, $\tilde{h} = h/L_D$, $\tilde{R} = r/L_D$ and time $\tilde{t} = t/t_e$. $L_D$ is a characteristic screening length.

Strong accumulation for both donors and electrons occurs in the corresponding regions adjacent to electrodes. However, as we can see from the comparison of **Figs 2a** and **2b** the maximal concentration of electrons is several times more than the maximal concentration of donors, despite the fact that initial concentration for both types of charge carriers was the same. Such situation results from the steric effect for donors, while electrons are regarded size-less. In the case of donor-blocking and electron-blocking electrodes, the total concentration of charge carriers in domain remains constant, while only local redistribution of carries concentration occurs. This corresponds to the periodical successive generation and annihilation of dynamic ionic-electron quasi-dipole and so to the formation and reorientation of corresponding polarization. For the case of the low voltage local extremums of the donor's concentration always appear on the same regions regardless of the sign of applied voltage, however it's not true for case of electrons, where number of local extremums are different for different time moments.

Redistribution of donors and electrons leads to emergence of stress and strain fields. Obtained elastic strains originate from donors motion. Radial ($\sigma_{rr}$) and azimuthal ($\sigma_{\varphi\varphi}$) components have similar distribution, which in turn highly correlate with donors concentration distribution (compare **Figs 2a,b** and **3a,b**). Normal stress component $\sigma_{zz}$, unlike the $\sigma_{rr}$ and $\sigma_{\varphi\varphi}$ components, has more uniform distribution. Generally, the strongest strain appears in the regions with the largest donors concentration gradients (directly under the probe). Comparison of the surface deformation in the **Fig.3a** and **3b, 3c** and **3d** for two different time moments $\tilde{t} = 10$, 20 illustrates the nonlinearity of the ESM response, indeed the absolute values of the local extremums are different during the first and second half period of the applied voltage. The first extremum (directly under the probe) almost doesn't change its absolute value and position; however the second extremum changes not only the absolute value, but also its position.

Electric potential monotonically increases with $\tilde{z}$, however for some moments ($\tilde{t} = 10$) local redistribution of charge carriers cause relatively small extremums. (see **Fig. 4a**). Also it was found out that changing of the $t_d/t_e$ relationship does not impact noticeably on the potential depth distribution. **Figure 4b** presents donors concentration $\tilde{z}$ – distribution. Naturally, under applying voltage to the top electrode donors move to the grounded electrode (see donor concentration in the **Fig. 4a)**. With increasing of the applied voltage amplitude, concentration of donors near the corresponding electrodes increases, reaching the values dozen times larger than the initial one. More interesting part is appearing of the local extremums of concentration, which can be interpreted as dynamic charge waves. As it can be seen from **Figure 4b** the donors charge wave in $\tilde{z}$-direction consists of two local extremums: one of them is local maximum and another one is the local minimum with several times higher absolute value. Note, that charge waves for the donors and electrons (shown in the Appendix B, Suppl. Mat.) have a different scale. Electron's charge waves are more flattened, their amplitude stays almost constant with changing of applied voltage and generally they are not so prominent in comparison with donor ones.

**Figures 5** shows that the structure of charge waves can be even more complicated in radial direction. In particular it demonstrates the appearance of the several local extremums. In general, donor's concentration dependence from radius is quite complicated, while for electrons such dependence has no more than one local extremum that moves toward the opposite electrode with almost constant amplitude. Note, that the computed dependence of concentration on the radial coordinate is not permanent across the domain volume. Namely, it changes drastically in the region adjacent to the bottom electrode. Here the concentration dependence from the radius for electrons no more have any local extremums, while the concentration for donors has only one extremum.

All figures are plotted for the ratio $t_d/t_e \equiv 10$. The same dependencies calculated for the $t_d/t_e = 10^2$ and $t_d/t_e = 10^3$ are shown in the **Appendix B, Suppl. Mat.** The general trend is that decreasing of $t_d/t_e$ parameter leads to a more pronounced local extremums for donors and almost do not impact on extremums for electron's dependencies. In more details crests become wider and have bigger amplitude; they are also shifted more further from the electrode. Increasing of the $t_d/t_e$ leads to opposite changes.

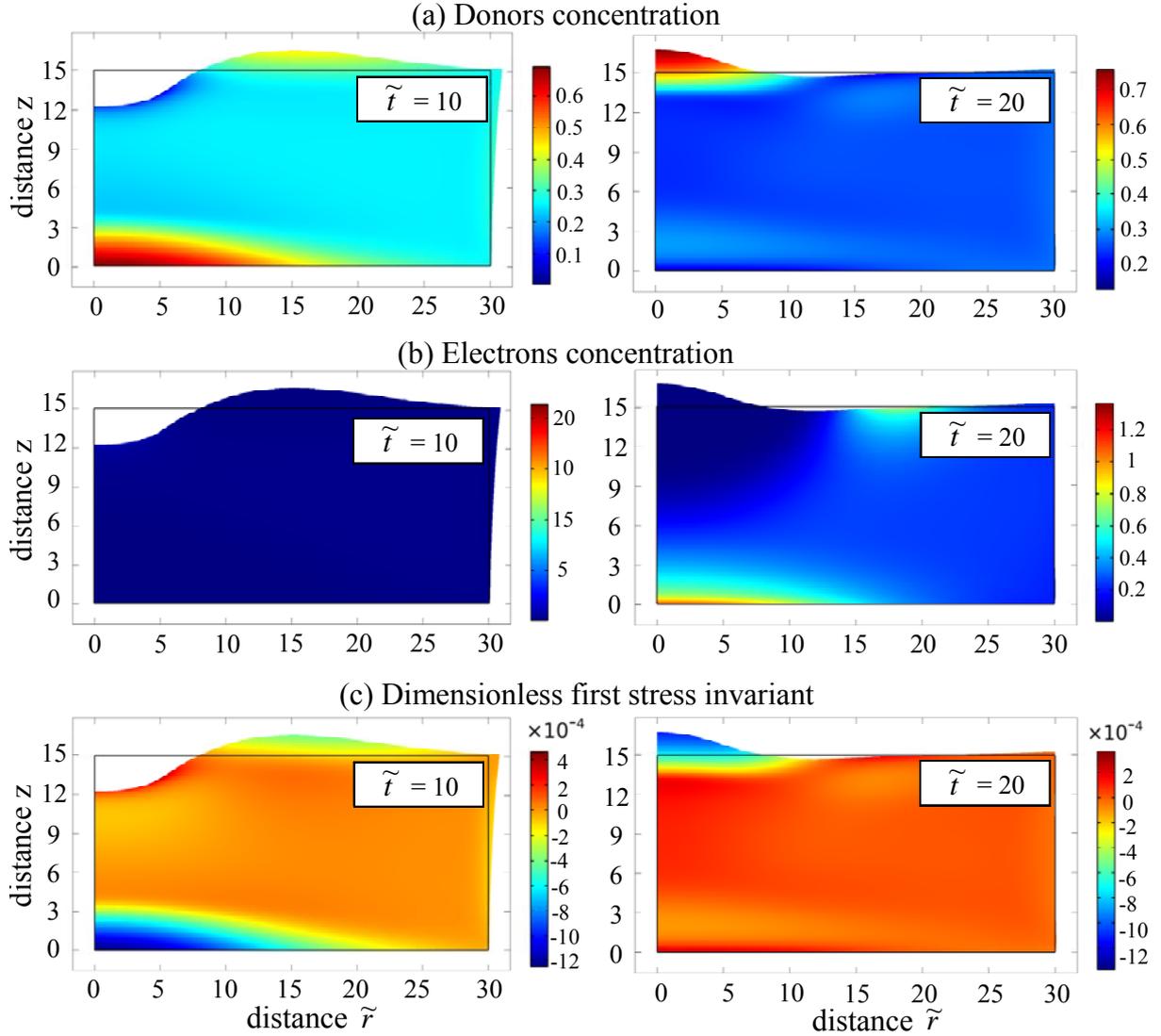

**Figure 2.** Dimensionless donors (**a**), electrons (**b**) concentration and the stress invariant $\sigma = \sigma_{zz} + \sigma_{rr} + \sigma_{\varphi\varphi}$ (**c**) for time moments $\tilde{t} = 10, 20$. The shape profile deviation from the empty rectangle shows its deformation scaled with factor $4 \cdot 10^3$. Due to the axial symmetry only the semi-slice of the cylinder is shown. Ratio $t_d/t_e \equiv 10$. Electrodes are donor- and electron-blocking. Applied voltage is 1V, $L_D = 2.8 \cdot 10^{-9} m$, $t_e = 8 \cdot 10^{-5} s$.

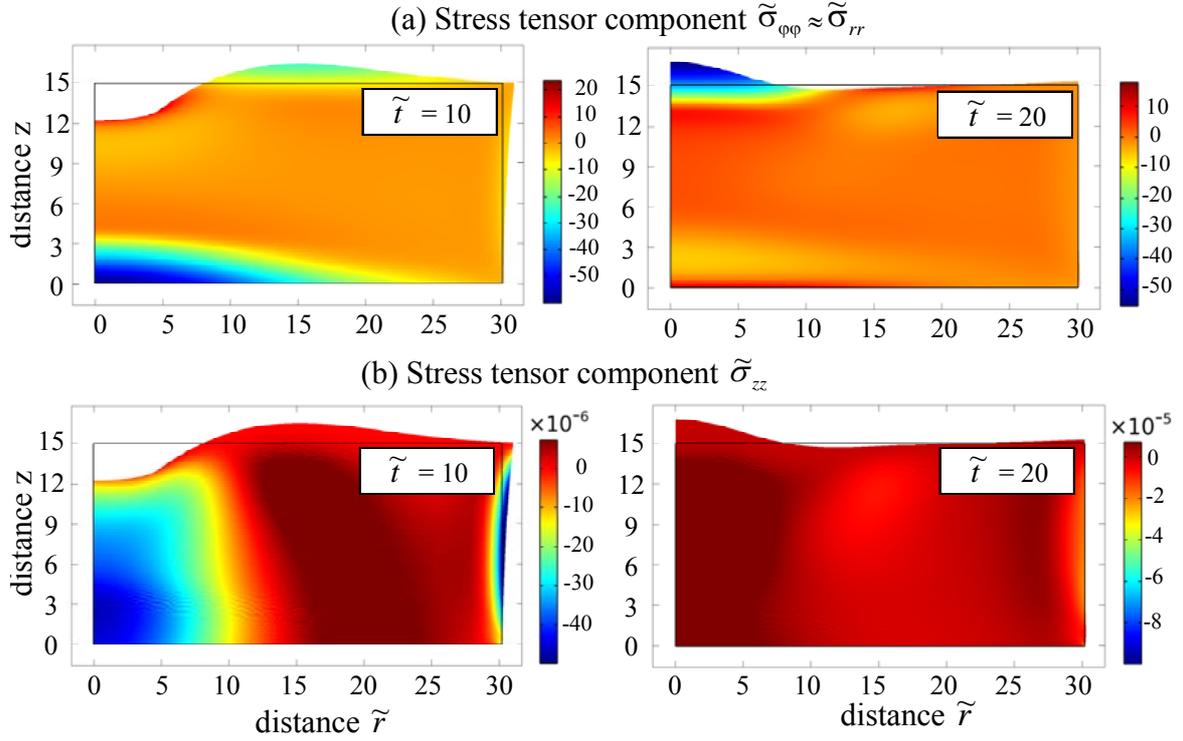

**Figure 3.** Dimensionless stress tensor $\tilde{\sigma}_{\varphi\varphi} \approx \tilde{\sigma}_{rr}$ (a) and $\tilde{\sigma}_{zz}$ (b) components for time moments $\tilde{t} = 10, 20$. Other parameters are the same as in the figure 2.

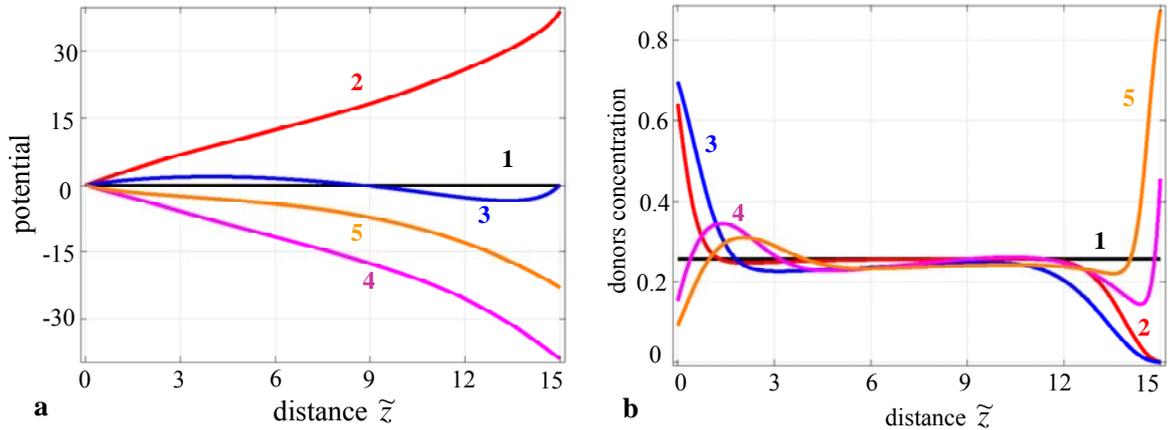

**Figure 4.** (a) Potential $\tilde{z}$ – distribution along the axes $\tilde{r} = 0$ and (b) donors concentration $\tilde{z}$ – distribution along the axes $\tilde{r} = 0$ calculated for donor-blocking and electron-blocking electrodes at different moments of time $\tilde{t} = 0, 5, 10, 15, 18$ (curves 1, 2, 3, 4 and 5).

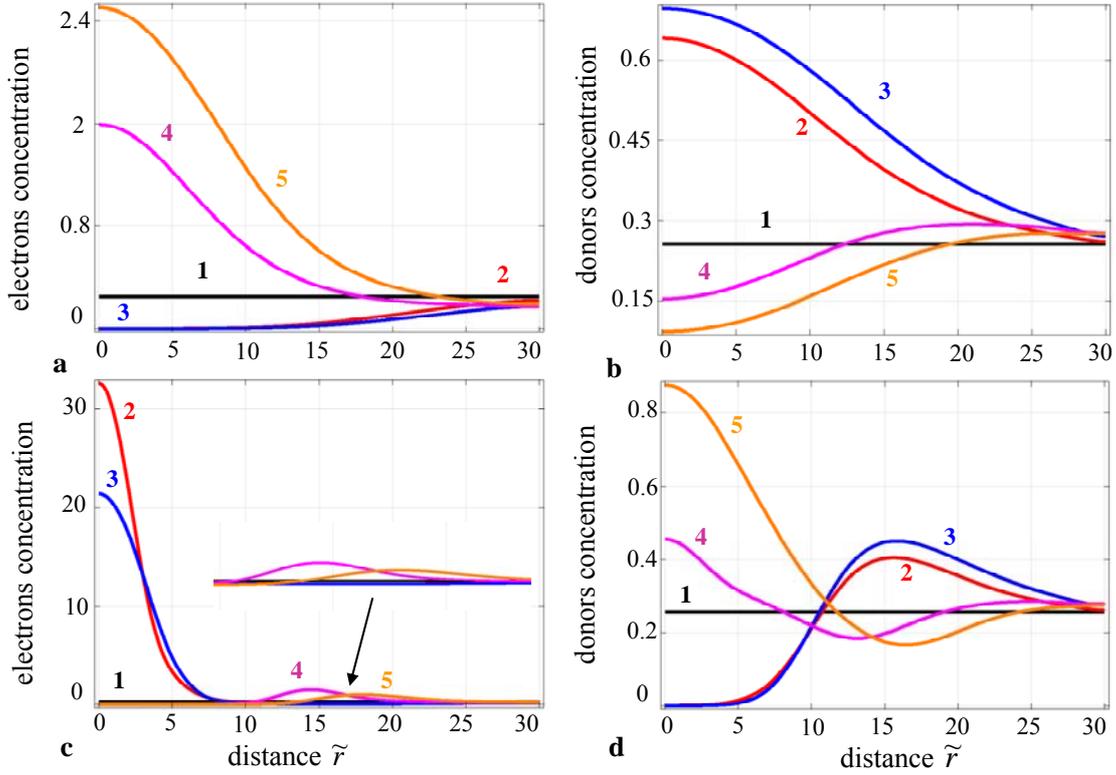

**Figure 5.** Electrons and donors $\tilde{r}-$ distribution along the axes $\tilde{z}=0$ **(a)**, **(b)** and $\tilde{z}=\tilde{h}$ **(c)**, **(d)** at different moments of time $\tilde{t}=0, 5, 10, 15, 18$ (curves 1, 2, 3, 4 and 5). Ratio $t_d/t_e \equiv 10$.

### 4. Summary

The self-consistent modeling of dynamic Electrochemical Strain Microscopy (ESM) for axially symmetric 2D mixed ionic-electronic conductors was performed in the COMSOL Multiphysics package. 2D map of the strain and concentration distribution across the sample and bias-induced surface displacements were obtained in the framework of the Thomas–Fermi screening theory and Vegard law with account of the steric effects for donors. Obtained results show significant impact of the nonlinear effects on the ESM image formation mechanisms (i.e. appearance of the charge waves and nonlinear deformation of the surface). It was found that Vegard mechanism plays a key role in the mechanisms of the ESM image formation at low voltages ≤1V. We realize that electrostriction contribution will dominate with voltage increase.

**Acknowledgements**

M.V S. acknowledges the grant of the President of the Russian Federation for state support of young Russian scientists-PhD (№14.125.13.3784) the project part of the State tasks in the field of scientific activity № 11.2551.2014/K. O.V.V., E.A.E. and A.N.M. acknowledge Center for Nanophase Materials Sciences, user projects CNMS 2013-293, CNMS 2014-270 and National Academy of Sciences of Ukraine (grant 35-02-14).

**References**

[1] Neeraj Panwar, I. K. Bdikin, A.N. Morozovska, and A. L. Kholkin. Domain growth kinetics in $La_{0.89}Sr_{0.11}MnO_3$ single crystal studied by piezoresponse force microscopy. J. Appl. Phys. **112**, 052019 (2012).

[2] Sergei V. Kalinin, Nava Setter, and Andrei L. Kholkin. "Electromechanics on the nanometer scale: emerging phenomena, devices, and applications." *MRS bulletin* 34, no. 09: 634-642 (2009).

[3] Alexander K. Tagantsev, Vincent Meunier, and Pradeep Sharma. "Novel electromechanical phenomena at the nanoscale: phenomenological theory and atomistic modeling." MRS bulletin 34, no. 09: 643-647 (2009).

[4] Dawn A. Bonnell, Sergei V. Kalinin, A. L. Kholkin, and A. Gruverman. "Piezoresponse force microscopy: a window into electromechanical behavior at the nanoscale." MRS bulletin 34, no. 09: 648-657 (2009).

[5] P. Muralt, R. G. Polcawich, and S. Trolier-McKinstry. "Piezoelectric thin films for sensors, actuators, and energy harvesting." MRS bulletin 34, no. 09: 658-664 (2009); F. Sachs, W. E. Brownell, and A. G. Petrov. "Membrane electromechanics in biology, with a focus on hearing." MRS bulletin 34, no. 09: 665-670 (2009).

[6] C. S. Meena, S. A. Mengi, and S. G. Deshpande, *Proc. Indian Acad. Sci. Chem. Sci.* **111**, 319 (1999).

[7] S. Ducharme, T. J. Reece, C. M. Othon, and R. K. Rannow, *IEEE Trans. Device Mater. Reliab.* **5**, 720 (2005).

[8] K. Kimura and H. Ohigashi, *Jpn. J. Appl. Phys.* **25**, 383 (1986).

[9] S. V. Kalinin, B. J. Rodriguez, S. Jesse, P. Maksymovych, K. Seal, M. Nikiforov, A. P. Baddorf, A. L. Kholkin, R. Proksch, *Materials Today* **11**, 16 (2008).

[10] A. Gruverman, A. Kholkin, Nanoscale ferroelectrics: processing, characterization and future trends. Rep. Prog. Phys. 69, 2443–2474 (2006).

[11] S. V. Kalinin, Anna N. Morozovska, Long Qing Chen, and Brian J. Rodriguez. "Local polarization dynamics in ferroelectric materials." *Reports on Progress in Physics* 73, no. 5: 056502(2010).

[12] A. Kumar, Oleg S. Ovchinnikov, H. Funakubo, S. Jesse, and S. V. Kalinin. "Real-space mapping of dynamic phenomena during hysteresis loop measurements: Dynamic switching spectroscopy piezoresponse force microscopy." *Applied Physics Letters* 98, no. 20: 202903 (2011).

[13] N. Balke, S. Jesse, A.N. Morozovska, E. Eliseev, D.W. Chung, Y. Kim, L. Adamczyk, R.E. Garcıa, N. Dudney and S.V. Kalinin. Nanoscale mapping of ion diffusion in a lithium-ion battery cathode. Nature Nanotechnology 5, 749–754. (2010).

[14] Amit Kumar, Francesco Ciucci, A.N. Morozovska, Sergei Kalinin, and Stephen Jesse, Measuring oxygen reduction/evolution reactions on the nanoscale. Nature Chemistry **3**, 707-713 (2011)

[15] Stephen Jesse, Nina Balke, Eugene Eliseev, Alexander Tselev, Nancy J. Dudney, Anna N. Morozovska, and Sergei V. Kalinin Direct Mapping of Ionic Transport in a Si Anode on the Nanoscale: Time Domain Electrochemical Strain Spectroscopy Study. ACS Nano. **5** (12), 9682–9695 (2011)

[16] N. Balke, E. A. Eliseev, S. Jesse, S. Kalnaus, C. Daniel, N. J. Dudney, A. N. Morozovska, and S. V. Kalinin, Three-dimensional vector electrochemical strain microscopy. J. Appl. Phys. **112**, 052020 (2012)

[17] Amit Kumar, Thomas M. Arruda, Yunseok Kim, Ilia N. Ivanov, Stephen Jesse, Chung W. Bark, Nicholas C. Bristowe et al. "Probing Surface and Bulk Electrochemical Processes on the LaAlO3–SrTiO3 Interface." *ACS nano* 6, no. 5: 3841-3852 (2012).

[18] Balke, Nina, Stephen Jesse, Yoongu Kim, Leslie Adamczyk, Ilia N. Ivanov, Nancy J. Dudney, and Sergei V. Kalinin. "Decoupling electrochemical reaction and diffusion processes in ionically-conductive solids on the nanometer scale." *ACS nano* 4, no. 12: 7349-7357 (2010).

[19] Nina Balke, Stephen Jesse, Yoongu Kim, Leslie Adamczyk, Alexander Tselev, Ilia N. Ivanov, Nancy J. Dudney, and Sergei V. Kalinin. "Real space mapping of Li-ion transport in amorphous Si anodes with nanometer resolution." *Nano letters* 10, no. 9: 3420-3425 (2010).

[20] S. R. Bishop, K. L. Duncan, and E. D. Wachsman. "Defect equilibria and chemical expansion in non-stoichiometric undoped and gadolinium-doped cerium oxide." *Electrochimica Acta* 54, no. 5: 1436-1443(2009).

[21] G. G. Botte, V. R. Subramanian, and R. E. White, *Electrochimica Acta*, **45**, 2595-2609 (2000).

[22] D.B.;Strukov, G.S. Snider, D. R. Stewart, R. S. Williams, The Missing Memristor Found. Nature 2008, 453, 80–83.

[23] K. Szot, W. Speier, G. Bihlmayer, R. Waser, Switching the Electrical Resistance of Individual Dislocations in Single-Crystalline SrTiO3. Nat. Mater., 5, 312 (2006).

[24] S. Jesse, H. N.Lee, & S. V. Kalinin, Quantitative mapping of switching behavior in piezoresponse force microscopy. Rev. Sci. Instrum. 77, 073702 (2006).

[25] A.N. Morozovska, E.A. Eliseev, A.K. Tagantsev, S.L. Bravina, Long-Qing Chen, and S.V. Kalinin. Phys. Rev. B 83, 195313 (2011).

[26] A.N. Morozovska, E.A. Eliseev, S.V. Kalinin. Appl. Phys. Lett. 96, 222906 (2010).

[27] A.N. Morozovska, E.A. Eliseev, N. Balke, S.V. Kalinin. J. Appl. Phys. 108, 053712 (2010).

[28] A.N. Morozovska, E.A. Eliseev, S.L. Bravina, Francesco Ciucci, G.S. Svechnikov, Long-Qing Chen and S.V. Kalinin. Frequency Dependent Dynamical Electromechanical Response of Mixed Ionic-Electronic Conductors. J. Appl. Phys. **111**, 014107 (2012)

[29] Mustafa Sabri Kilic, Martin Z. Bazant, and Armand Ajdari. *Physical Review* **E 75**, 021502 (2007)

[30] Mustafa Sabri Kilic, Martin Z. Bazant, and Armand Ajdari. *Physical Review* **E 75**, 021503 (2007)

[31] Anna N. Morozovska, Eugene A. Eliseev, P.S.Sankara Rama Krishnan, Alexander Tselev, Evgheny Strelkov, Albina Borisevich, Olexander V. Varenyk, Nicola V. Morozovsky, Paul Munroe, Sergei V. Kalinin and Valanoor Nagarajan. Defect thermodynamics and kinetics in thin strained ferroelectric films: the interplay of possible mechanisms. Phys.Rev.B 89, 054102 (2014).

[32] Daniel A. Freedman, D. Roundy, and T. A. Arias, Phys. Rev. B 80, 064108 (2009).

[33] X. Zhang, A. M. Sastry, W. Shyy, J. Electrochem. Soc. **155**, A542 (2008).

[34] H.-Ch. Chang and G. Jaffe, J. Chem. Phys. **20**, 1071 (1952).

[35] Supplementary materials.

# Supplemental Materials to
# Self-consistent modelling of nonlinear dynamic ESM microscopy in mixed ionic-electronic conductors

O.V. Varenyk[1], M. Silibin[2], D. Kiselev[2,3], E.A. Eliseev[4], S.V. Kalinin[5][1], A.N. Morozovska[1][2]

[1] *Institute of Physics, National Academy of Sciences of Ukraine, 46, pr. Nauki, 03028 Kyiv, Ukraine*

[2] *National Research University of Electronic Technology "MIET", 124498 Moscow, Russia*

[3] *National University of Science and Technology "MISiS", 119049 Moscow, Leninskiy pr. 4, Russia*

[4] *Institute for Problems of Materials Science, NAS of Ukraine, Krjijanovskogo 3, 03142 Kiev, Ukraine*

[5] *The Center for Nanophase Materials Sciences, Oak Ridge National Laboratory, Oak Ridge, TN 37831*

---

[1] Corresponding author 1
[2] Corresponding author 2

## Appendix A. Electrostriction contribution

For the case of the mixed ionic-semiconductor film placed in a planar capacitor under the voltage $V_{dc} + V_{ac}\sin(\omega t)$, electric potential $\varphi$ can be found self-consistently from the Poisson equation with the short-circuited electric boundary conditions

For the case electric potential $\varphi$ can be found self-consistently from the Poisson equation with the short-circuited electric boundary conditions:

$$\begin{cases} \varepsilon_0 \varepsilon_{33}^b \dfrac{\partial^2 \varphi}{\partial z^2} = -\rho(\varphi), \\ \varphi(0) = V_{dc} + V_{ac}\sin(\omega t), \quad \varphi(h) = 0. \end{cases} \qquad (A.1)$$

Here the frequency $\omega$ is much smaller than the optical frequency and we regard that $|V_{dc}| \gg |V_{ac}|$. Under the assumption:

$$\varphi(z) = \varphi_{dc} + \varphi_{ac}\sin(\omega t), \qquad (A.2a)$$

$$\rho(\varphi) = \rho_{dc} + \rho_{ac}\sin(\omega t) \qquad (A.2b)$$

Note that for typical ionic systems the mobilities are sufficiently slow so that at ~100 kHz excitation frequencies the $E_{ac}$ component can be evaluated as that for dielectric, neglecting ionic motion. Below we will use the ***depletion/accumulation layer approximation*** for the determination of dc components and ***Debye (or Tomas-Fermi) approximation*** for the determination of ac components. Namely:

$$\rho_{dc} \approx \begin{cases} \rho_0, & 0 < z < l \\ 0, & l < z < h \end{cases} \qquad (A.3a)$$

$$\rho_{ac} \approx \dfrac{\varphi_{ac}}{h_d^2} \qquad (A.3b)$$

The value $\rho_0$ is determined by equilibrium concentrations of holes and electrons in the quasi-neutral region of the semiconductor [[i]]. Debye screening length is $h_d$. Note that for high frequencies $\omega \gg 1/\tau_M$ (where $\tau_M$ is the Maxwellian relaxation time) the density $\rho_{ac}$ is almost zero due to the very sluggish ionic dynamics and so the electric field amplitude $E_{ac} = V_{ac}/h$.

Quasistatic ***dc electric potential and field*** are:

$$\varphi_{dc}(z) = -\dfrac{\rho_0 (l-z)^2}{2\varepsilon_0 (1+\chi)}\theta(h-z), \qquad E_3^{dc}(z) = \dfrac{\rho_0 (l-z)}{2\varepsilon_0 (1+\chi)}\theta(l-z) \qquad (A.4)$$

Here $\theta(z)$ is the step-function. Relative static lattice relative permittivity is $(1+\chi)$. The thickness $h$ should be found self-consistently from the boundary condition $\varphi(0) = V_{dc}$, that gives

$$l = \sqrt{2\varepsilon_0(1+\chi)V_{dc}/\rho_0} \tag{A.5}$$

The semiconductor potential and space charge are distributed in the layer $0 < z < h$ and zero outside. Trivially, the limits of the depletion layer approximation validity is the condition $l \ll h$, the limit is $\rho_0 h^2 > 2\varepsilon_0(1+\chi)V_{dc}$. It is very important limitation; it makes impossible to consider a continuous transition to the dielectric limit. However, the rigorous solution using Eq.(1) will be free from the limitation.

Dynamic *ac electric potential and field* calculated in Debye approximation is

$$\varphi_{ac} = -V_{ac}\frac{\sinh((z-h)/h_d)}{\sinh(h/h_d)}, \qquad E_3^{ac} = \frac{V_{ac}}{h_d}\frac{\cosh((z-h)/h_d)}{\sinh(h/h_d)} \tag{A.6}$$

Note that Eqs.(6) contains a continuous transition to the dielectric limit, $E_{ac} = V_{ac}/h$, at $h_d \to \infty$ that occurs at $\omega \gg 1/\tau_M$. Small $h_d \ll h$ corresponds to the carriers instant response to the ac field.

Finally we can estimate the film strain and surface displacement caused by electrostriction in linear dielectric approximation ($P_3 = \varepsilon_0\chi E_3$):

$$u_3^Q(t) = Q_{33}(\varepsilon_0\chi)^2\int_0^L \left(E_3^{dc}(z) + E_3^{ac}\sin(\omega_0 t)\right)^2 dz \approx u_3^0 + u_3^\omega\sin(\omega_0 t) + u_3^{2\omega}\cos(2\omega_0 t) \tag{A.7}$$

The coefficients

$$u_3^0 = Q_{33}(\varepsilon_0\chi)^2\frac{\rho_0^{1/2}V_{dc}^{3/2}}{3\sqrt{2\varepsilon_0(1+\chi)}}, \tag{A.8a}$$

$$u_3^\omega = \frac{Q_{33}(\varepsilon_0\chi)^2\rho_0 V_{ac}}{2\varepsilon_0(1+\chi)}\left(\frac{h_d\cosh\left((h-\sqrt{2\varepsilon_0(1+\chi)(V_{dc}/\rho_0)})/h_d\right)}{\sinh(h/h_d)} - h_d\coth(h/h_d) + \sqrt{2\varepsilon_0(1+\chi)(V_{dc}/\rho_0)}\right) \xrightarrow[h_d\to\infty]{} Q_{33}(\varepsilon_0\chi)^2\frac{V_{dc}V_{ac}}{h}, \tag{A.8b}$$

$$u_3^{2\omega} = -Q_{33}(\varepsilon_0\chi)^2\left(\frac{V_{ac}}{2h_d}\right)^2\frac{2L + h_d\sinh(2h/h_d)}{2\sinh^2(h/h_d)} \xrightarrow[h_d\to\infty]{} -\frac{1}{2}Q_{33}(\varepsilon_0\chi)^2\left(\frac{V_{ac}}{h}\right)^2. \tag{A.8c}$$

Finally, for the cases of linear Debye screening or no screening of *ac* component and abrupt junction approximation for *dc* component, the total displacement becomes:

$$u_3^Q(t) = u_3^0 + u_3^\omega\sin(\omega t) + u_3^{2\omega}\cos(2\omega t) \tag{A.9}$$

Coefficients $u_3^{2\omega} = -Q_{33}(\varepsilon_0\chi)^2(V_{ac}/h)^2/2$, $u_3^\omega = 2Q_{33}\varepsilon_0\chi P_W V_{ac} + Q_{33}(\varepsilon_0\chi)^2(V_{dc}V_{ac})/h$ and $u_3^0 = Q_{33}P_W^2 h + Q_{33}\varepsilon_0\chi P_W V_{dc} + Q_{33}(\varepsilon_0\chi)^2(\rho_0^{1/2}V_{dc}^{3/2})/3\sqrt{2\varepsilon_0(1+\chi)}$ have different powers on $V_{dc}$ and $V_{ac}$. The nontrivial behavior of occurs $u_3^0$ from non-linear screening phenomena. Static lattice relative permittivity is $(1+\chi)$.

Since we are dealing with cylindrical coordinate system, $E_3 = 0$, hence $u_{33} = u_{32} = u_{31} = u_{23} = u_{13} = 0$, $Q_{11} = 0.11$ m$^4$/C$^2$, $Q_{12} = -0.045$ m$^4$/C$^2$, $\chi_{11} = \chi_{22} = 5$, $\varepsilon_0 = 8.85 \times 10^{-12}$ F/m, $u_{12} = \varepsilon_0^2 \frac{Q_{11} - Q_{12}}{2} \chi_{11} \chi_{22} \frac{(k_B T)^2}{e^2 L_D^2} \frac{\partial^2 \tilde{\varphi}}{\partial \tilde{x}_1 \partial \tilde{x}_2}$. Numerical estimation for $\varepsilon_0^2 \frac{Q_{11} - Q_{12}}{2} \chi_{11} \chi_{22} \frac{(k_B T)^2}{e^2 L_D^2} = 12.4 \times 10^{-9}$, So we are getting quite predictable result of small electrostriction impact as shown in the **Figure A1**.

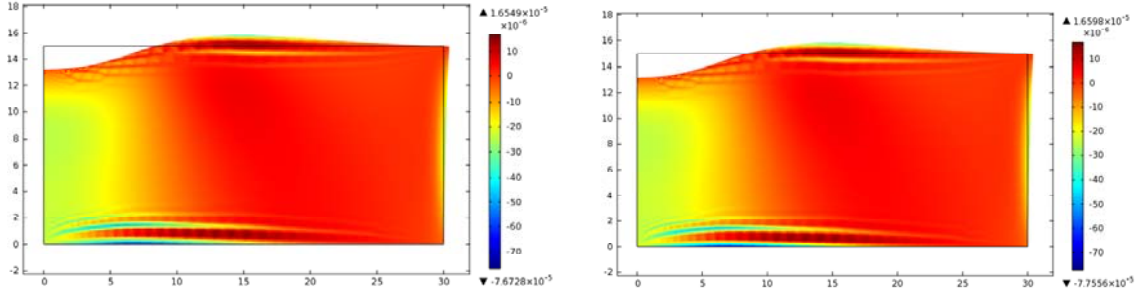

**Figure A1.** Dimensionless stress tensor z component with electrostriction is left figure, without is right figure

## Appendix B. Dimensionless equations

In dimensionless variables, the equations for chemical potentials (8)-(9) coupled with Poisson equation for electric potential (1), Lame's equation (12) and corresponding boundary conditions are listed below.

$$\frac{t_d}{t_e} \frac{\partial (f(\tilde{\mu}_d - \tilde{E}_d))}{\partial \tilde{t}} - \frac{1}{\tilde{r}} \frac{\partial}{\partial \tilde{r}} \left( \tilde{r} f(\tilde{\mu}_d - \tilde{E}_d) \frac{\partial}{\partial \tilde{r}} (\tilde{\varphi} - \tilde{\mu}_d + \tilde{w}^2 f(\tilde{\mu}_d - \tilde{E}_d)) \right) - \frac{\partial}{\partial \tilde{z}} \left( f(\tilde{\mu}_d - \tilde{E}_d) \frac{\partial}{\partial \tilde{z}} (\tilde{\varphi} - \tilde{\mu}_d + \tilde{w}^2 f(\tilde{\mu}_d - \tilde{E}_d)) \right) = 0$$
(B.1a)

$$\frac{\partial}{\partial \tilde{t}} \left( F_{1/2}(\tilde{\mu}_e - \tilde{E}_C) \right) - \frac{1}{\tilde{r}} \frac{\partial}{\partial \tilde{r}} \left( \tilde{r} F_{1/2}(\tilde{\mu}_e - \tilde{E}_C) \frac{\partial}{\partial \tilde{r}} (\tilde{\mu}_e - \tilde{\varphi}) \right) - \frac{\partial}{\partial \tilde{z}} \left( F_{1/2}(\tilde{\mu}_e - \tilde{E}_C) \frac{\partial}{\partial \tilde{z}} (\tilde{\mu}_e - \tilde{\varphi}) \right) = 0$$
(B.1b)

$$\frac{\partial^2 \tilde{\varphi}}{\partial \tilde{r}^2} + \frac{1}{\tilde{r}} \frac{\partial \tilde{\varphi}}{\partial \tilde{r}} + \frac{\partial^2 \tilde{\varphi}}{\partial \tilde{z}^2} = -\left( \frac{N_d^0}{N_d^+} f(\tilde{\mu}_d - \tilde{E}_d) - \frac{N_C}{N_d^+} F_{1/2}(\tilde{\mu}_e - \tilde{E}_C) \right)$$
(B.1c)

$$\tilde{c}_{ijkl} \frac{\partial^2 \tilde{u}_k}{\partial \tilde{x}_j \partial \tilde{x}_l} - \tilde{c}_{ijkl} \cdot \tilde{\beta}_{kl} \frac{\partial \delta \tilde{N}_d^+}{\partial \tilde{x}_j} = 0$$
(B.1d)

$$\tilde{J}_d \Big|_{\tilde{z}=0} = 0; \quad \tilde{J}_d \Big|_{\tilde{z}=\tilde{h}} = 0; \quad \tilde{J}_d \Big|_{\tilde{r}=\tilde{R}} = 0$$
(B.2a)

$$\left( \tilde{J}_e - \tilde{\xi}_0 (\tilde{n} - \tilde{n}_b) \right) \Big|_{\tilde{z}=0} = 0, \quad \left( \tilde{J}_e + \tilde{\xi}_h (\tilde{n} - \tilde{n}_b) \right) \Big|_{\tilde{z}=\tilde{h}} = 0, \quad \tilde{J}_e \Big|_{\tilde{r}=\tilde{R}} = 0$$
(B.2b)

$$\tilde{\varphi} \Big|_{\tilde{z}=0} = \tilde{V} \sin(\tilde{\omega} \tilde{t}), \quad \tilde{\varphi} \Big|_{\tilde{z}=\tilde{h}} = 0, \quad \tilde{\varphi} \Big|_{\tilde{r}=\tilde{R}} = 0.$$
(B.2c)

$$\left( \tilde{c}_{ijkl} \frac{\partial \tilde{u}_k}{\partial \tilde{x}_l} - \tilde{c}_{ijkk} \tilde{W}_{kl} \, \delta \tilde{N}_d^+ \right) n_j \bigg|_{\tilde{z}=\tilde{h}} = 0 \,, \quad \tilde{u}_k \big|_{\tilde{z}=\tilde{h}} = 0 \qquad (B.2d)$$

**Table B1. Dimensionless variables and parameters**

| Quantity | Definition/designation |
|---|---|
| Dimensionless coordinate and thickness | $\tilde{z} = z/L_D$, $\tilde{r} = r/L_D$, $\tilde{h} = h/L_D$, $\tilde{R} = r/L_D$ |
| Debye screening length | $L_D = \sqrt{\varepsilon_0 \varepsilon_{33}^b k_B T / (e^2 \overline{N}_d^+)}$ |
| Dimensionless time | $\tilde{t} = t/t_e$ |
| Dimensionless frequency of applied voltage | $f = t_e \omega / 2\pi$ |
| Characteristic electronic and donor times | $t_e = L_D^2 / (e \eta_e k_B T)$, $t_d = L_D^2 / (e \eta_d k_B T)$ |
| Dimensionless donor concentration | $\tilde{N} = N_d^+ / \overline{N}_d^+ = (N_d^0 / \overline{N}_d^+) f(\tilde{\mu}_d - \tilde{E}_d)$ |
| Dimensionless electron density | $\tilde{n} = n / \overline{N}_d^+ = (N_C / \overline{N}_d^+) F_{1/2}(\tilde{\mu}_e - \tilde{E}_C)$ |
| Equilibrium concentration of ionized donors at zero potential and stress | $\overline{N}_d^+ = N_d^0 (1 - f((E_d - E_F)/k_B T))$ |
| Effective density of states in the conductive band | $N_C = \left( \dfrac{m_n k_B T}{2\pi \hbar^2} \right)^{3/2}$ |
| Dimensionless electric potential | $\tilde{\varphi} = e \varphi / k_B T$ |
| Applied voltage | $\tilde{V} = eU / k_B T$ |
| Dimensionless chemical potentials | $\tilde{\mu}_d = \mu_d / k_B T$, $\tilde{\mu}_e = \mu_e / k_B T$ |
| Dimensionless donor level | $\tilde{E}_d = E_d / k_B T$ |
| Dimensionless conduction band position | $\tilde{E}_C = E_C / k_B T$ |
| Dimensionless Fermi energy | $\tilde{E}_F = E_F / k_B T$ is determined self-consistently from the electroneutrality $N_C F_{1/2}(\tilde{E}_F - \tilde{E}_C) = N_d^0 f(\tilde{E}_F - \tilde{E}_d)$ |
| Dimensionless Vegard coefficient | $\tilde{w}^2 = \dfrac{2W^2 \overline{N}_d^+}{(s_{11} + s_{12}) k_B T}$ |
| Dimensionless electron and donor currents | $\tilde{J}_e = \dfrac{L_D J_e}{e \overline{N}_d^+ \eta_e k_B T}$, $\tilde{J}_d = \dfrac{L_D J_d}{e \overline{N}_d^+ \eta_e k_B T}$, |
| Total electric current is the sum of electronic, donor and displacement components | $\tilde{J} = \tilde{J}_e + \tilde{J}_d + \partial \tilde{E} / \partial \tilde{t}$ |
| Dimensionless rate constant and concentration | $\tilde{\xi}_{d,e} = \dfrac{L_D v_{d,e}}{e \eta_e k_B T}$, $\tilde{n}_b = \dfrac{n_b}{\overline{N}_d^+}$ |
| Tensor of elastic stiffness | $\tilde{c}_{ijkl} = c_{ijkl} / Y$ |
| Vegard expansion tensor | $\tilde{W}_{kl} = W_{kl} \overline{N}_d^+$ |
| Mechanical displacement | $\tilde{u}_k = u_k / L_D$ |

**Table B2. Values of parameters used in the numerical calculations**

| Parameter or ratio | Numerical values and comments |
|---|---|
| $\tilde{h}$ | 15 |
| $t_d/t_e \equiv \eta_e/\eta_d$ | 103, $10^2$, 10 |
| $f$ | 1/20 |
| $E_C$ (eV) | 0 |
| $E_d$ (eV) | −0.1 |
| $E_F$ (eV) | Determined self-consistently from the electroneutrality condition |
| $m_n/m_0$ | 0.5 |
| $W$ (Å$^3$) | 10 |
| $s_{11}+s_{12}$ (Pa$^{-1}$) | $3.44 \times 10^{-12}$ |
| $N_d^0$ (m$^{-3}$) | $10^{24}$ |
| $T$ (K) | 298 |
| $\varepsilon$ | 5 |
| $\tilde{n}_b$ | 1 |
| $\tilde{\xi}_{0,h}$ | 0 |
| $D_i$ | $10^{-14}$ |
| $D_e$ | $10^{-12}$ |
| $\eta_e$ | $2.43 \times 10^8$ |
| $\eta_i$ | $2.43 \times 10^6$ |
| $\tilde{w}^2$ | 0.01413 |
| $L_D$ | $2.841 \times 10^{-9}$ |
| $N_C$ | $4.39 \times 10^{24}$ |
| $\tilde{Y}$ | 1 |
| $\tilde{W}_{kl}$ | $1 \cdot 10^{-3}$ |

**Figure B1** plot **a** shows potential distribution along r direction with constant coordinate $z = \tilde{h}$. Due to the symmetry of applied voltage function, some curves completely coincide, therefore there are only five different curves on the plot. Generally potential is a monotonic function from $\tilde{r}$ coordinate which exponentially increases by absolute value with approaching domain center and saturates to constant value in the near center region. In $\tilde{z}$ direction with constant coordinate $\tilde{r} = 0$, potential primarily monotonically increases with increasing of $\tilde{z}$, however for some moments ($\tilde{t} = 10$) local redistribution of charge carriers cause relatively small extremums. Also it was find out that changing of the $t_d/t_e$ relationship almost do not impact on potential distribution across domain.

**Figure B2** presents electrons and donors concentration $\tilde{z}$ – distribution. One can note that with applying voltage to the top electrode the electrons move to the positive electrode while donors move to the ground one. With increasing of the applied voltage amplitude, concentration of both electrons and donors near the corresponding electrodes increases, reaching the values dozen times larger for the initial. More interesting part it is appearing of the local extremums of concentration which can be interpreted as dynamic charge waves. While moving toward the opposite electrodes, the amplitude of the crests changes with exponential like law. As it can be seen from **Figure B2b** such charge wave for donors, in $\tilde{z}$ -direction consists of two local extremums: one of them is local maximum and another one is the local minimum with several times bigger absolute value. Comparison of the **Figure B2** plot **a** and plot **b** show that charge waves for the donors and electrons have a different scale. Electron's charge waves are more flattened, their amplitude stays almost constant with changing of applied voltage and generally they are not so prominent compare to donor's one.

**Figure B3** and **Figure B4** show the same dependencies for donors calculated for the $t_d/t_e = 10^2$ and $t_d/t_e = 10^3$ respectively. The general trend is that decreasing of $t_d/t_e$ parameter leads to a more pronounced local extremums for donors and almost do not impact on extremums for electron's dependencies. In more details crests become wider and have bigger amplitude (near 1.5 times for $t_d/t_e = 10$ compare to $t_d/t_e = 10^2$ ). They are also shifted more further from the electrode. Increasing of the $t_d/t_e$ parameter leads to opposite changes. **Figure B2** plot **a**, **b** and **c** shows that crests are very narrow and slightly change their position.

**Figures B5** and **B6** show that in radial direction structure of charge waves can be even more complicated. For instance we demonstrate the appearance of the three local extremums. In general, donor's concentration dependence from radius is quite complicated, while for electrons such dependence has no more than one local extremum which moves toward the opposite electrode with almost constant amplitude.

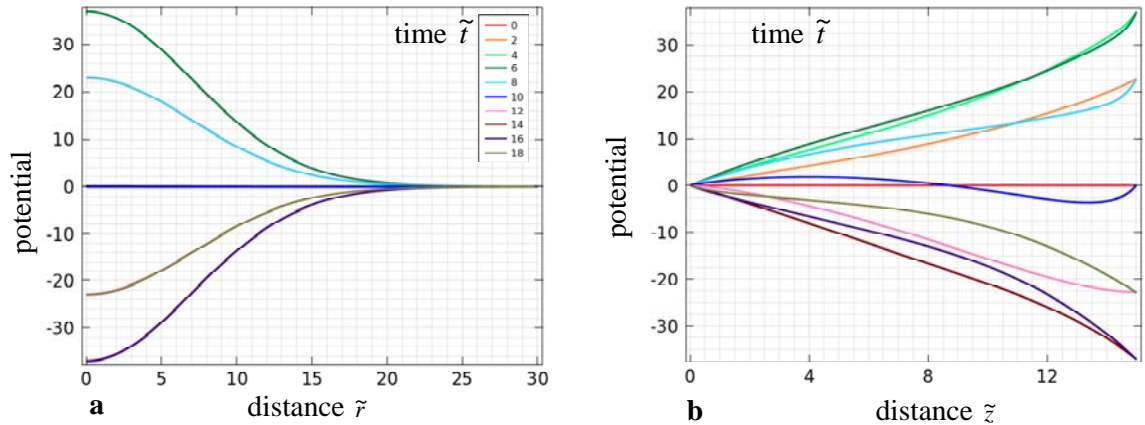

**Figure B1.** (a) Potential $\tilde{r}$ − distribution along the axes $\tilde{z} = \tilde{h}$ and (b) $\tilde{z}$ − distribution along the axes $\tilde{r} = 0$ calculated for donor-blocking and electron-blocking electrodes at different moments of time. Ratio $t_d/t_e \equiv 10$

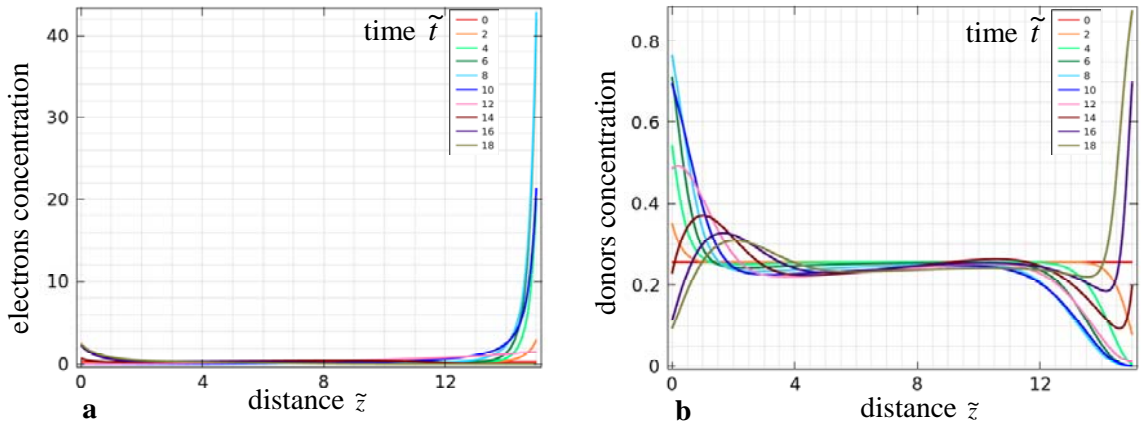

**Figure B2.** Electrons (**a**), donors (**b**) concentration $\tilde{z}$ – distribution along the axes $\tilde{r} = 0$ calculated for donor-blocking and electron-blocking electrodes at different moments of time. Ratio $t_d/t_e \equiv 10$

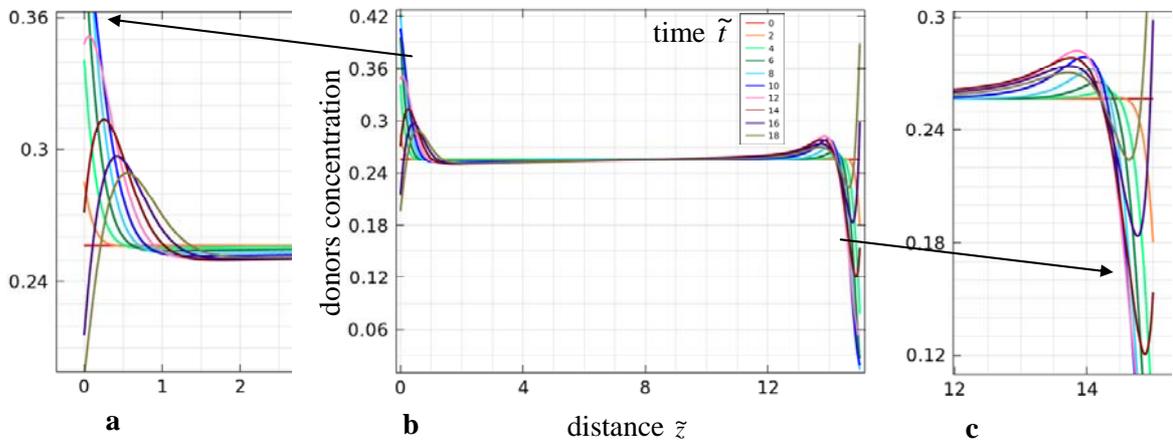

**Figure B3.** Donors (**b**) concentration $\tilde{z}$ – distribution along the axes $\tilde{r} = 0$ calculated for donor-blocking and electron-blocking electrodes at different moments of time. Ratio $t_d/t_e \equiv 10^2$

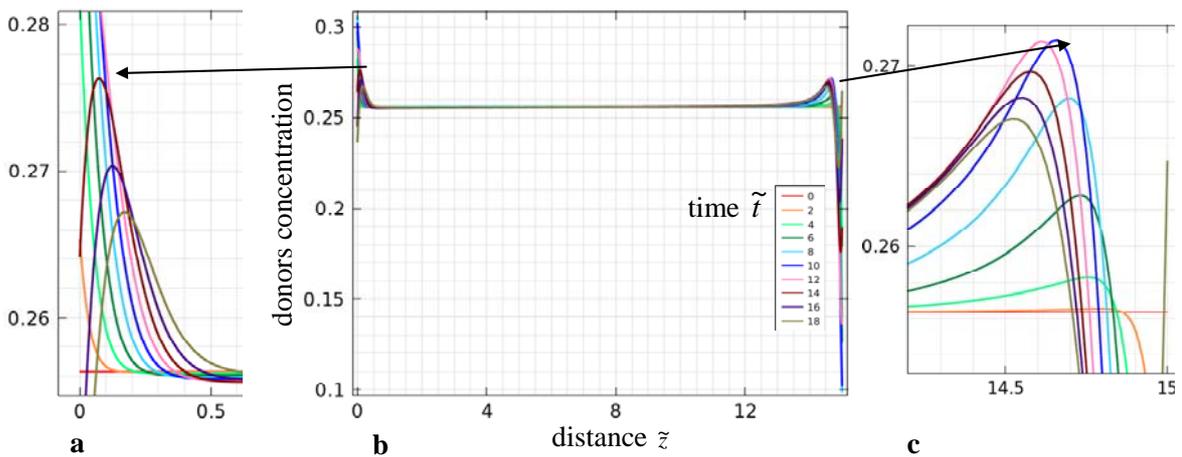

**Figure B4.** Donors $\tilde{z}$ – distribution along the axes $\tilde{r} = 0$ calculated for donor-blocking and electron-blocking electrodes at different moments of time. Ratio $t_d/t_e \equiv 10^3$

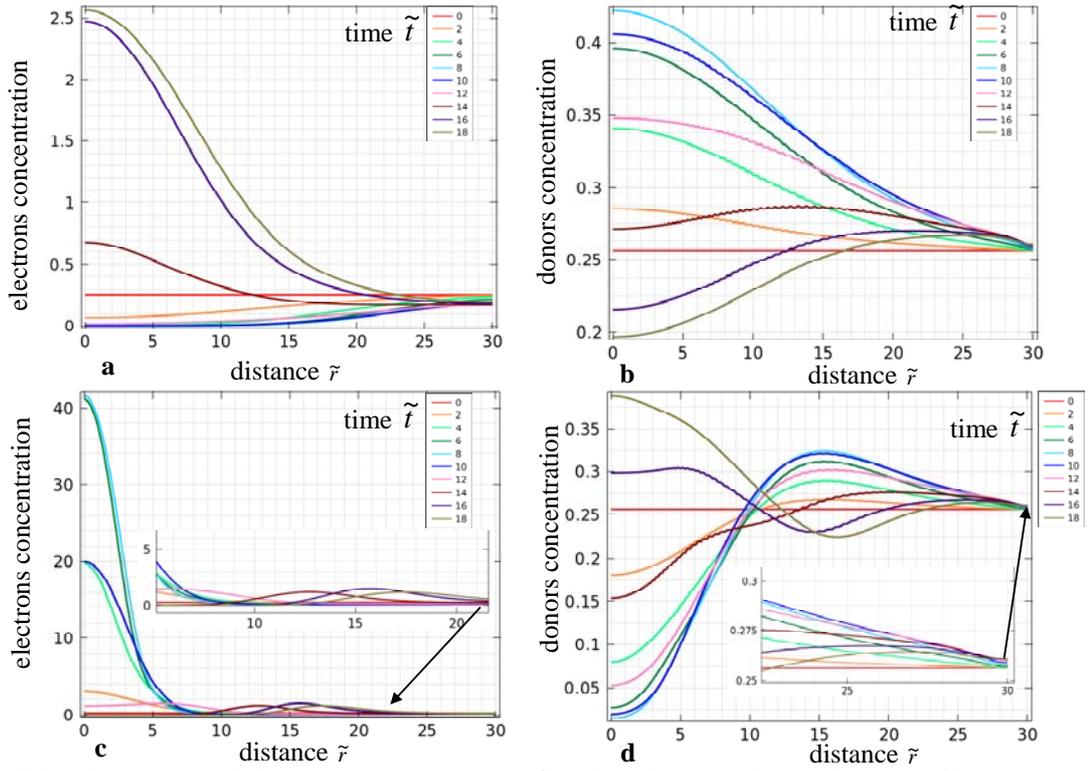

**Figure B5.** Electrons and donors concentration $\tilde{r}$ – distribution along the axes $\tilde{z}=0$ **(a)**, **(b)** and $\tilde{z}=\tilde{h}$ **(c)**, **(d)** calculated for donor-blocking and electron-blocking electrodes at different moments of time. Ratio $t_d/t_e \equiv 10^2$

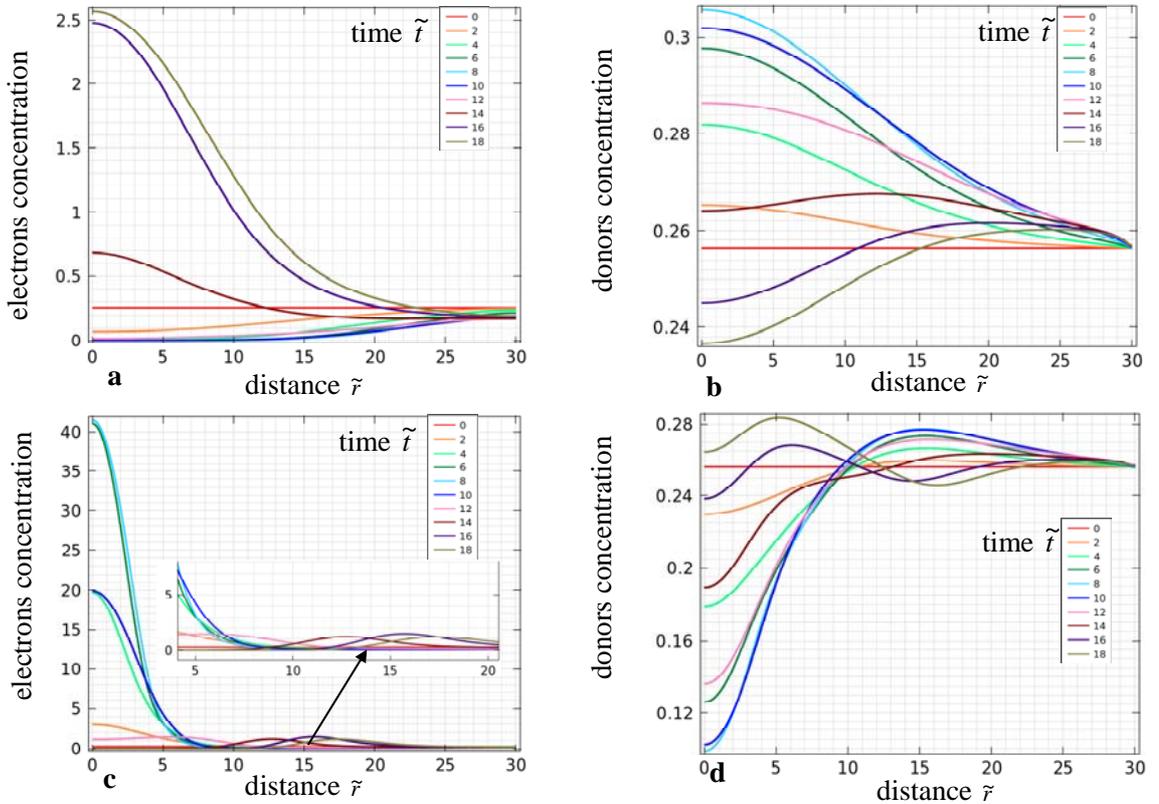

**Figure B6.** Electrons and donors $\tilde{r}$ – distribution along the axes $\tilde{z}=0$ **(a)**, **(b)** and $\tilde{z}=h$ **(c)**, **(d)** calculated for donor-blocking and electron-blocking electrodes at different moments of time. Ratio $t_d/t_e \equiv 10^3$

[i] S. M. Sze, *Physics of Semiconductor Devices*, 2nd ed. (Wiley-Interscience, New York, 1981), Chap. 7, p. 382.